\documentstyle[11pt,aaspp4]{article}

\tighten

\slugcomment{submitted to ApJ.}

\newcommand{\etal}{{\it et al.\ }}
\newcommand{\xiav}{\bar{\xi}}

\newcommand{\ltsima}{$\; \buildrel < \over \sim \;$}
\newcommand{\lsim}{\lower.5ex\hbox{\ltsima}}
\newcommand{\gtsima}{$\; \buildrel > \over \sim \;$}
\newcommand{\gsim}{\lower.5ex\hbox{\gtsima}}

\begin{document}

\title{A New Method for Calculating\\
       Counts in Cells}

\author{Istv\'an Szapudi\altaffilmark{1}}
\affil{University of Durham, Department of Physics}
\affil{South Road, Durham, DH1 3LE, United Kingdom}

\altaffiltext{1}{E-mail:\ Istvan.Szapudi@durham.ac.uk}

\begin{abstract}
In the near future a new generation of CCD based galaxy
surveys will enable high precision determination of
the $N$-point correlation functions. The resulting 
information will help to resolve the ambiguities associated
with two-point correlation functions thus constraining
theories of structure formation, biasing, and Gaussianity of initial 
conditions independently of the value of $\Omega$. As one the most successful
methods to extract the amplitude of higher order correlations is based on
measuring the distribution of counts in cells, this work
presents an advanced way of measuring it
with unprecedented accuracy. 
Szapudi and Colombi (1996, hereafter~\cite{sc96}) identified the main
sources of theoretical errors in extracting
counts in cells from galaxy catalogs. 
One of these sources, termed as measurement error,
stems from the fact that
conventional methods use a finite number of sampling
cells to estimate counts in cells. This effect can
be circumvented by using an infinite number of cells.
This paper presents an algorithm, which, {\it in practice} achieves
this goal, i.e. it is equivalent to throwing an infinite
number of sampling cells in finite time.  
The errors associated with sampling cells are completely eliminated
by this procedure which will be essential for 
the accurate analysis of future surveys.

\end{abstract}


\pagebreak

\section{Introduction}

Since the two-point correlation function and its Fourier counter
part, the power spectrum does not contain phase information,
higher order statistics are needed for full description of
the (highly non-Gaussian) galaxy density field. The 
immediate generalization of the
two-point correlation function is the set of $N$-point
correlation functions, which corresponds to higher order
moments of a spatial distribution.  These objects are, however, 
difficult to measure and interpret mainly because of the
combinatorial explosion of terms and the small configuration
space associated with them, especially for high orders. Therefore
other (indirect) methods, such as  moments of counts in cells,
bi-counts in cells, void probability, structure functions 
on minimal spanning trees, wavelet methods, genus, etc. became
popular alternatives. Of these, counts in cells techniques
(\cite{peebles80,gaz92,bouchet93,gaz94,cbh95}, Szapudi, Meiksin \&
Nichol 1996, hereafter \cite{smn96}) were
some of the most successful in the recent past; the aim of this
paper is to present a new technique for calculating counts in cells
with unprecedented accuracy. First, however, to substantiate the need of such
a method, their connection to $N$-point correlation functions is 
reviewed briefly. 

Mathematically counts in cells and $N$-point correlation functions
are equivalent according to well known theorems in spatial statistics,
which state that, if suitable conditions hold,
 factorial moment measures and random measures
on Borel sets are equivalent description of a discrete random
process (see e.g. Ripley 1988). 
While the former are related to the $N$-point correlation
functions, the latter are essentially counts in cells.
In practice, however, one never extracts all possible configurations, and
to infinite order (higher order correlation functions), and neither
uses all possible cell shapes and sizes (counts in cells), so in both
cases information is lost. Therefore the two approaches are somewhat
complementary in their information content, 
and there is also a difference in efficiency and accuracy. 
The choice between them is of entirely practical nature;
generally both are interesting to study.

Direct determination of the $N$-point correlation functions
using $DD\ldots D$ type methods can in principle measure the
full $N$-point function, while counts in cells method only gives
its smoothed version in a cell. Therefore the former is inherently
more accurate than the latter, if the sole purpose is to
extract higher order correlation functions.
However, measurement of $N$-point
correlation functions is burdened with the large number of variables
they depend on. The CPU time becomes
exponentially expensive with order ($\simeq N_{par}^{N}$,
where $N_{par}$ is the number of galaxies), and the
results are extremely difficult to interpret, as they
depend on many variables ($3N-5$). In contrast, counts in
cells are easy to measure, once
they are measured, any order can be straightforwardly calculated,
and the results can be easily interpreted, as in the
simplest cases the only parameter is the scale of the cell.
On the other hand, configuration dependence can be studied by using
a series of different shapes for sampling cells.
The errors on counts in cells measurements are
smaller for high orders than the corresponding direct
determination because of larger the configuration
space available for averaging. Presently, full non-linear analytical
formulae are available for the errors on moments of
 counts in cells including all higher order effects (\cite{sc96}),
while only cruder approximations are possible for the $N$-point
functions.

The most successful version of  the
technique calculates the factorial moments 
and cumulants (\cite{ss93a}, see also \cite{bs89}), 
from the distribution of galaxy counts in cells. The resulting cumulants,
or amplitudes of the higher order correlation functions according
to the definition
$S_N = \xiav_N/\xiav_2^{N-1}$, in turn can be compared with results from
perturbation theory (\cite{peebles80,jbc93,bern92,bern94}), 
$N$-body simulations, and the theory
gravitational statistics based on the BBKGY equations
(\cite{dp77,peebles80,cbh95,bge95,sqsl97}). These theories
assuming gravity and Gaussian initial conditions predict
a certain set of cumulants, $S_N$'s, while non-Gaussian initial
conditions (\cite{col92}), and biasing (\cite{fg94}) have
different predictions. Therefore high precision determination
of the $S_N$'s in fully sampled CCD based catalogs, such as the 
future SDSS, will be crucial in resolving the ambiguities associated
with the two-point correlation functions (and its reincarnations)
to constrain theories of structure formation, biasing, and
the nature of initial conditions. 

Note that similar statements are true about
cumulant correlators (\cite{ss97a}),
a matrix version of the $S_N$'s which is based on
bi-counts in cells.
These contain more information although 
slightly more complicated
than regular counts in cells, 
but are simpler to calculate,
although carry somewhat less configuration information 
than the full $N$-point functions.

The above arguments explain why
direct measurement of the higher order correlation functions
(ex. \cite{peebles80}) is complicated for $N > 4$, and
accurate methods based on
counts in cells became crucial for understanding higher order statistics
of the distribution of galaxies. However, especially with new powerful
computers and new methods designed to eliminate edge effects
(\cite{ss97b}), extracting $N$-point functions will gain more interest
in the near future. 
Moreover, for low order moments,
such as $N \le 4$ direct methods are certainly viable, and as outlined
above, contain more accurate shape information. 
Despite these recent and anticipated advances,
 there will always be a degree of complementarity 
between direct determination of $N$-point functions,
and counts in cells, thus advanced measurement techniques are
most useful for both.
This is the motivation
for the method presented here to extract counts in cells
with unprecedented accuracy by diminishing the errors associated
with sampling cells.

~\cite{sc96} examined in detail the
problem of errors on statistics related to counts in cells.
They found, that theoretical errors fall into two distinct
classes: cosmic errors (including finite volume effects,
discreteness effects, and edge effects), and measurement
errors. While the former is an inherent property of the
galaxy catalog at hand, thus can be improved upon only
by creating a larger, denser catalog, the second one can be eliminated
in principle by throwing an infinite number of cells. As discussed
in~\cite{sc96}, the number of cells one needs to throw 
(``number of independent cells'') depends on the statistic
and scale at question. 
The asymptotic behavior of the errors is proportional $1/C$, where $C$ is the
number of sampling cells, with the constant of proportionality increasing
toward higher order quantities
and smaller scales. 
While at least massive oversampling is recommended
to control the errors up to a certain order,
only infinite sampling makes the measurement error term
completely disappear for all order. Surprisingly, 
infinite sampling can be achieved in practice.
This work presents such method
with moderate CPU investment compared to the alternative 
of mending the traditional procedure with massive oversampling.
The next section describes the algorithm,
in \S3 evaluates a practical implementation, presents
measurements, and discusses the relevance of the results.

\section{The Algorithm}

The basic observation underlying the method is that
the measurement of  counts in cells by throwing 
an infinite number of random cells is equivalent to
a series of integrals over step functions.  
These can be evaluated to arbitrary precision
without actually throwing {\it any} cells.
Thus the traditional way of throwing random cells corresponds
to a Monte Carlo integration, while the
other popular method involving a grid 
is equivalent to Euler's formula. 
Here the exact calculation is proposed for ultimate accuracy.

Let me define the following set of functions
\begin{equation}
   f_N(x) = \left\{ \begin{array}{ll}
                      1 & \mbox{if $M = N$} \\
                      0 & \mbox{otherwise}
                    \end{array}
            \right.
\end{equation}
where $M$ is the number of objects within a cell centered
on $x$. Clearly the estimator for $P_N$ is
\begin{equation}
  P_N \simeq \lim_{C\rightarrow -> \infty}\frac{1}{C} \sum f_N(x_i)
  = \int_V d^3 x f_N(x), 
\end{equation}
where $C$ the number of random cells at positions $x_i$ tends
to infinity, and the Monte Carlo realization of the integral
approaches the integral itself. Obviously, calculating the
integral is equivalent to throwing an infinite number of sampling
cells. Exact calculation  is possible because the function $f_N$ is 
piecewise constant. Note also that $\sum f_N(x) = 1$ for any $x$, therefore
only one of the $f_N$'s can be non-zero. Also, for any finite
galaxy catalog, there exists a maximum number in the galaxy cell
counts (for instance it is bounded by the total number of
objects). These two properties
facilitate the computation of all the $\int f_N$'s simultaneously.

A geometric interpretation of
the above idea is most useful to devise an algorithm to calculate
the needed integrals exactly.
Figure 1. illustrates the problem of measuring counts in cells
for a special configuration.
There are four points in a rectangular box. 
Around each object (large dots) a square is drawn, identical to
the sampling cell used for counts in cells.
The possible
centers of random cells all lie within the dashed line, which
follows the boundary of the bounding box. 
Since the square around each point corresponds to the possible centers
of (random) cells containing  that same point, 
the question can be reformulated in the following way:
let us partition the area of the possible centers of cells
according to the overlap properties of the cells drawn
around the objects. If $N$ squares overlap
in a partition, then $f_N = 1$ throughout the partition, and the
rest of the $f_j$'s are all zero. This is illustrated with
different shadings on the figure. Thus the problem of calculating
the integral exactly is equivalent to finding the sum of areas
in the partitions for each $N$.

The above considerations, although illustrated with square cells,
apply to any cell shape, and for finite number of points. However, it
is easiest to determine overlaps of rectangular cells (in any
dimension), therefore the rest of the paper will be restricted
to rectangular shape. This is not a serious restriction, because
the shape dependence is not expected to be severe in the galaxy
distribution, even though spherical cells do have some theoretical
advantage such as being directionless.

One obvious possibility for calculating the needed overlaps
is a tree data structure (similar to a tree $N$-body
code) to find all the neighbors of a point for determining the
overlaps in an adaptive mesh. 
I found, however, that the 'sweep' paradigm from computational geometry 
can be used to construct a simpler and more memory efficient algorithm.
This can also be thought of as an adaptive grid covering the total area,
however, only the part immediately needed for the calculation is stored
in memory. 
For simplicity, I refer to the configuration on Figure 1.
in the following description of the method.
The calculation for any configuration should be obvious from this.

Imagine a rigid vertical line moving slowly from the
left of Figure 1. towards the right;
the boundary can be ignored temporarily.
Before the line touches any of the squares, it sweeps
through an area with $f_0 = 1$.
Therefore at the point of first contact all the swept
area contributes to $\int f_0$ and can be recorded.
 After the contact the line is divided
into segments sweeping through areas with $f_0 = 1$ and $f_1 = 1$
respectively. The boundaries of these segments can be imagined
as two markers on the line, corresponding to the upper and lower
corner the square being touched. As the sweep continues,
the results can be recorded at any contact with the 
side of a square during the movement of the line:
the areas swept are assigned according to the markers
on the line to different $\int f_N$'s.
This is done with a one dimensional sweep on the line  
counting the two kinds of
markers. Then the segmentation of the line is updated. Whenever the
line makes contact with the left side of a square, two markers are added,
whenever
it touches the right hand side of a square, the corresponding markers
are dropped. 
The boundaries and rectangular masks,
can be trivially taken into account by only starting to record 
the result of the sweep when entering the area of possible centers.
Non-rectangular masks can be converted to rectangular by putting
them on a grid.

If there are $N$ objects in the plane, the above procedure will
finish after $2 N$ updating.
The algorithm can be trivially generalized for arbitrary rectangles, 
any dimensions. For instance in three dimensions the basic
sweep is done with a plane, while
the plane has to be swept by a line  after each contact. 
The generalization for
circles, and spheres, or arbitrary shapes, seems to be fairly
complicated, although it might be possible.

\section{Discussion}

From the definition of the algorithm it follows that
the required CPU time scales as $N^D (d/L)^{D(D-1)/2}$ in
$D$ dimensions, where N is the number of objects, $d/L$ is
the ratio of the scale of measurement to the characteristic survey length.
Artificial galaxy catalogs were generated using {\tt ran1} from
\cite{numrec} in a rectangle of $19$ by $55$ degrees, matching
exactly the dimensions of the EDSGC catalog as used by \cite{smn96}.
Figure 2. shows the scaling measured for a 
family of two-dimensional catalogs.
The dashed line shows the approximate scaling
$t \simeq 2.8\ 10^{-8} N^2 d_{deg}$ on both panels,
which is in good agreement with the expectations. 
The memory requirement is approximately linear with $N$.

The accuracy of the code can be judged by inspecting
Figure 3. where a series of measurements are shown in 
a two-dimensional artificial catalog with a million
objects in it. The theoretical Poisson distribution
is shown with dotted, the infinitely sampled measurements
with solid lines.
The different curves correspond to a series of scales
ranging from $0.016$ to $2$ degrees.
The theoretical and measured curves agree perfectly
with each other. With massive oversampling, roughly
$10^8\ldots 10^{10}$ random cells would achieve the same accuracy.
Note, that Poisson distribution is actually simpler
to measure accurately than the long tailed distribution
of the galaxy surveys because of the non-Gaussian error
distribution (\cite{sc96}).

The code was also applied to real galaxy data (\cite{smn96}). 
On their Fig. 1.  the traditional method
of calculating counts in cells on a single grid totally misses
the shape of the probability distribution.
It was found that the 
infinite oversampling provided by the
proposed algorithm was most essential on small scales, where
Poisson noise can dominate the signal. 
In this regime undersampling
can severely underestimate the moments of the distribution, 
especially for higher order. This effect can be understood 
in terms of the theoretical results 
by \cite{sc96}, where the ``number of statistically
independent cells'' was found to increase sharply toward
smaller scales, and increasing order. Since the
error distribution is fairly skewed, 
from an ensemble of low sampled measurements many
will underestimate the moments, while a few will overestimate them
substantially. The sum will still give the right ensemble average 
identical to the infinitely oversampled measurements. 
This means that a particular undersampled measurement is likely
to underestimate the moments since the small number
of sampling cells can miss a rare cluster with high probability. 
Similarly, there is
a small chance of largely overestimating the moments
when, with a small probability, a cell happens to hit a rare cluster exactly.
In effect, this phenomenon can cause the unbiased statistical
estimator to give lower values for the moments. Only massive
oversampling, and preferably, the algorithm outlined in this
work can yield accurate, unbiased measurements.

Note, that ``infinite sampling'' is not strictly true, even
in the case of the algorithm presented here: in principle
one should sample all possible orientations of a square.
However, it was found theoretically, that the form
factors associated with square and circular cells
are quite similar (\cite{gaz94,bss94,cbh95}),
which was confirmed with measurements in simulations
(\cite{bge95}). Nevertheless, it is worth to note,
that if rectangular cells are used, with long aspect
ratios, especially in smaller survey, possibly dominated
with filamentary structures, it could be necessary
to sample more than one orientation. In most cases,
however, the proposed algorithm is close to 
``infinite sampling'' even with one orientation. To demonstrate
this, Figure 4. displays counts in cells measurements in 
the EDSGC survey with 1:2 cell aspect ratio. Detailed
description of the data can be found in \cite{smn96}. Note,
that the curvature of the sky was compensated with a projection
to physical (equal area) coordinates. Again, this is a good
approximation in light of the weak shape dependence.
The lower panel shows the raw counts in cells results:
there is a slight variation in the tail of the distribution,
but it is by no means significant. The upper panel shows this
even more clearly, where the higher order cumulants, $S_N$'s,
are displayed (see \cite{smn96} for the method
of the calculation).
The scale of the measurement is the side of an
equivalent square, i.e. geometrical average of the two sides.
The squares and stars show the two perpendicular
orientation measured, and up to 9th order they perfectly overlap.
The accuracy of the agreement can be judged from a few
sample values of fractional differences defined
as $S_N^{1:2}/S_N^{2:1}-1$. The largest deviation is
at $0.044$ degrees, and for $N = 3,4,5$, it is
$0.04, 0.21, 0.55$, respectively. For other scales
the fractional difference is typically an order of magnitude
less, except maybe for the largest scales, where it is only
a factor of two less than the quoted maximum values. 
For orientation, the original measurement by \cite{smn96} is
shown with triangles and solid line. Interestingly, the 
shape dependence seems to be negligible at most scales:
the largest deviation again occurs at $0.044$ degrees, but it
is still a lot smaller than the errors of the original measurement
(see \cite{smn96} for details). The Figure illustrates
that i) the proposed method can be used
in most cases without sampling different orientations
ii) the algorithm is capable of extracting shape dependence
of higher order statistics by using different rectangular
shapes. Exploring different cell shapes will in fact be useful
to disentangle biasing from gravitational amplification
({\cite{fss97}) in the near future. Note that while the technique proposed
here cannot be easily extended to bi-counts in cells
(e.g., \cite{ss97a}), it can be used to obtain similar
information by studying different aspect ratios.

As expected from the construction of the sweep, 
the CPU time for the real data of \cite{smn96} was of the same
order as for an artificial  catalog with same number of objects in it. 
The CPU time comparison with
the alternative of throwing {\it large} number of random cells is ambiguous, 
since the effective
number of sampling cells for the method of this work is infinity. 
On the data set of \cite{smn96} the
number of cells were increased in the traditional algorithm
using multiple oversampling grids until the resulting
irreducible $N$th moments do not change significantly.
It was found that order of twenty times more CPU was appropriate for
up to $9$th order. However, the results
of the infinite precision calculation are not only
faster, but more accurate as well. The convergence of
actually throwing a large number of cells is slow
because of the $1/C$ asymptotic.

While the above detailed tests were performed for the
two-dimensional version of the code, a three dimensional
version was implemented as well. Because
of the sharp increase in CPU time, proportional to $N^3$, 
this version is practical only for a moderate red shift survey
of tens of thousands of galaxies with widely available
computers. Perhaps supercomputers can remedy the situation
somewhat, since the algorithm is naturally parallelizable
via domain decomposition.
For $N$-body simulations
containing millions of particles, a pair of 
new algorithms will
be described elsewhere (\cite{sqsl97}). 

This paper presented a new method for the measurement
of counts in cells, a quantity central to higher order statistics.
The new method is equivalent to throwing an infinite number
of sampling cells in a traditional algorithm, and as such
eliminates the contribution to the 
``measurement errors'' (\cite{sc96}). This way the full
$1$ point information is extracted from the data if 
the negligible effect of sampling different orientations
is disregarded. 
The implementation
of the code is significantly more accurate, and orders of magnitude
faster than the traditional approach, making it a natural choice
for analyzing future galaxy surveys.

\acknowledgements

It is a pleasure to acknowledge discussions 
with S. Colombi, which motivated the need of a method
described in the present paper. I would like to thank
A. Szalay for discussions, and A. Stebbins for reading
the manuscript. 
I.S. was supported by DOE and NASA through grant
NAG-5-2788 at Fermilab.

\section{Figure Captions}

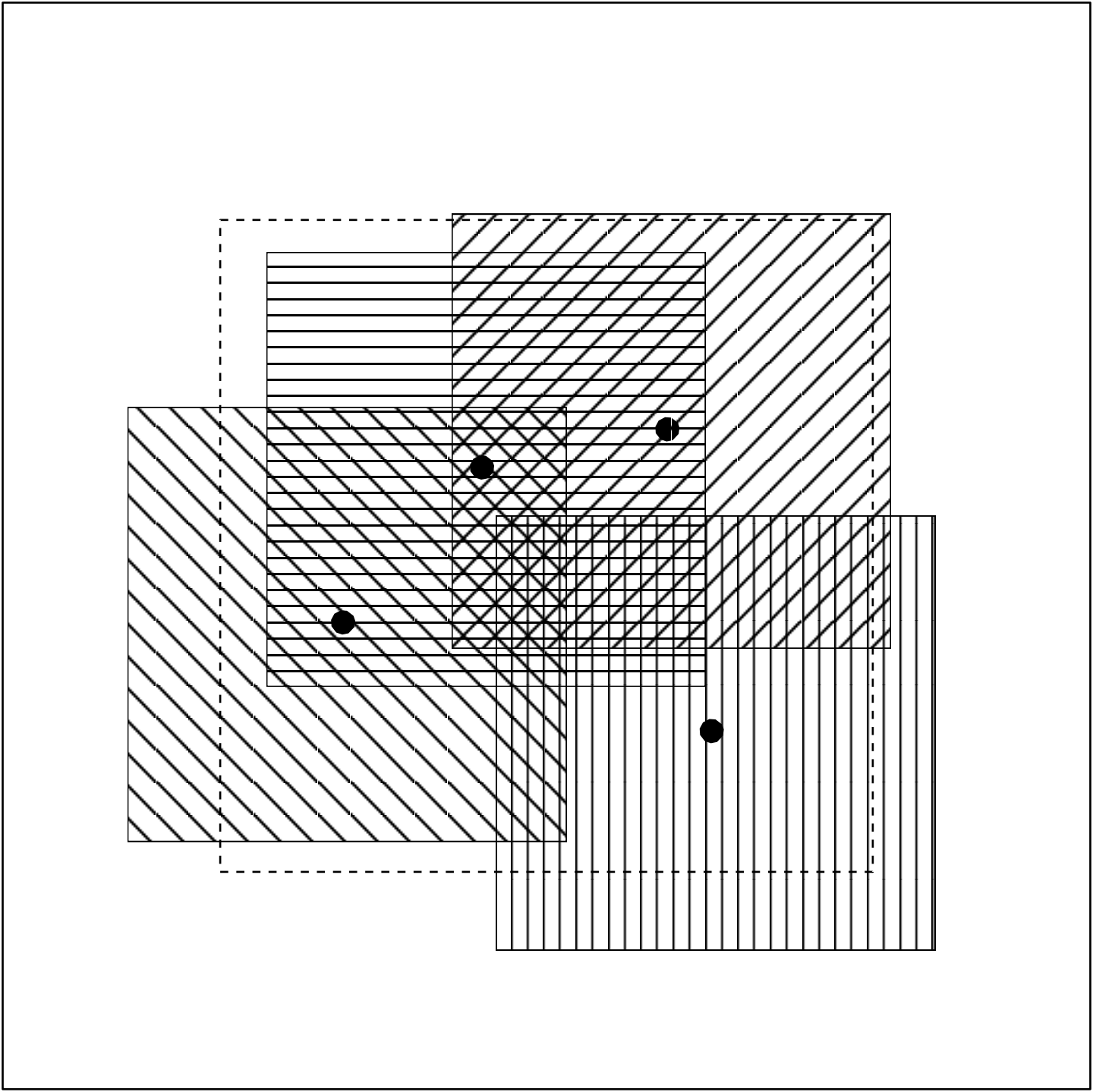
\noindent Figure 1.\ 
Illustrates the geometric calculation of counts in cells.
There are four points within the solid boundary. The centers
of square cells can lie within the dashed boundary. Around
each point a square is drawn to represent the possible centers
of cells which contain that point. The problem of counts
in cells can now be reformulated as calculation of the ratios
of all overlap areas (represented with different shadings
on the figure) within the dashed boundary. 

\noindent Figure 2.\ The CPU time of the measurements
of counts in cells in artificial galaxy catalogs is displayed.
The solid line represents the actual measurements, while
the dotted line is the theoretical scaling, 
$t \simeq 2.8\ 10^{-8} N^2 d_{deg}$, where the universal
constant was ``fit'' by a few trial.  Panel a.\ displays
the time as a function of the number of galaxies in the survey, 
while $d_{deg}$ is a parameter, doubling from $0.016$ to $2$
degrees from below. Panel b. \ displays $t$ is a function of $d_{deg}$,
while $N$ is $5\ 10^4, 10^5, 2\ 10^5, 2.9\ 10^5, 4\ 10^5$,
and $10^6$ from below.

\noindent Figure 3.\ Shows the measurement of counts in cells
in an artificial galaxy catalog of $19$ by $55$ degrees with
$N = 10^6$ galaxies. The measurements are shown with solid lines,
while the dotted lines display the theoretical curves. The agreement
shows the unprecedented accuracy of the proposed method.

\noindent Figure 4.\ Displays the measurement of counts in cells
in the EDSGC catalog with cell aspect ratio 1:2. The lower panel
shows the raw counts in cells results, with two orientations
displayed. The scales associated with the curves from left
to right are $0.015625\times 0.03125$ doubling to
$1\times 2$ degrees. The variance of the two curves is small,
and only present in the tails of the distribution.
The upper panel shows the $S_N$'s (obtained similarly to
\cite{smn96}) as a function of scale (the geometric mean
of the two sides of the rectangular sampling cell).
The squares and stars represent the two orientations:
they overlap perfectly. For orientation, the results
from \cite{smn96} are displayed as well with triangles
and solid lines. Note the excellent agreement, which
indicates a fairly weak shape dependence, and virtually
no dependence on orientation.

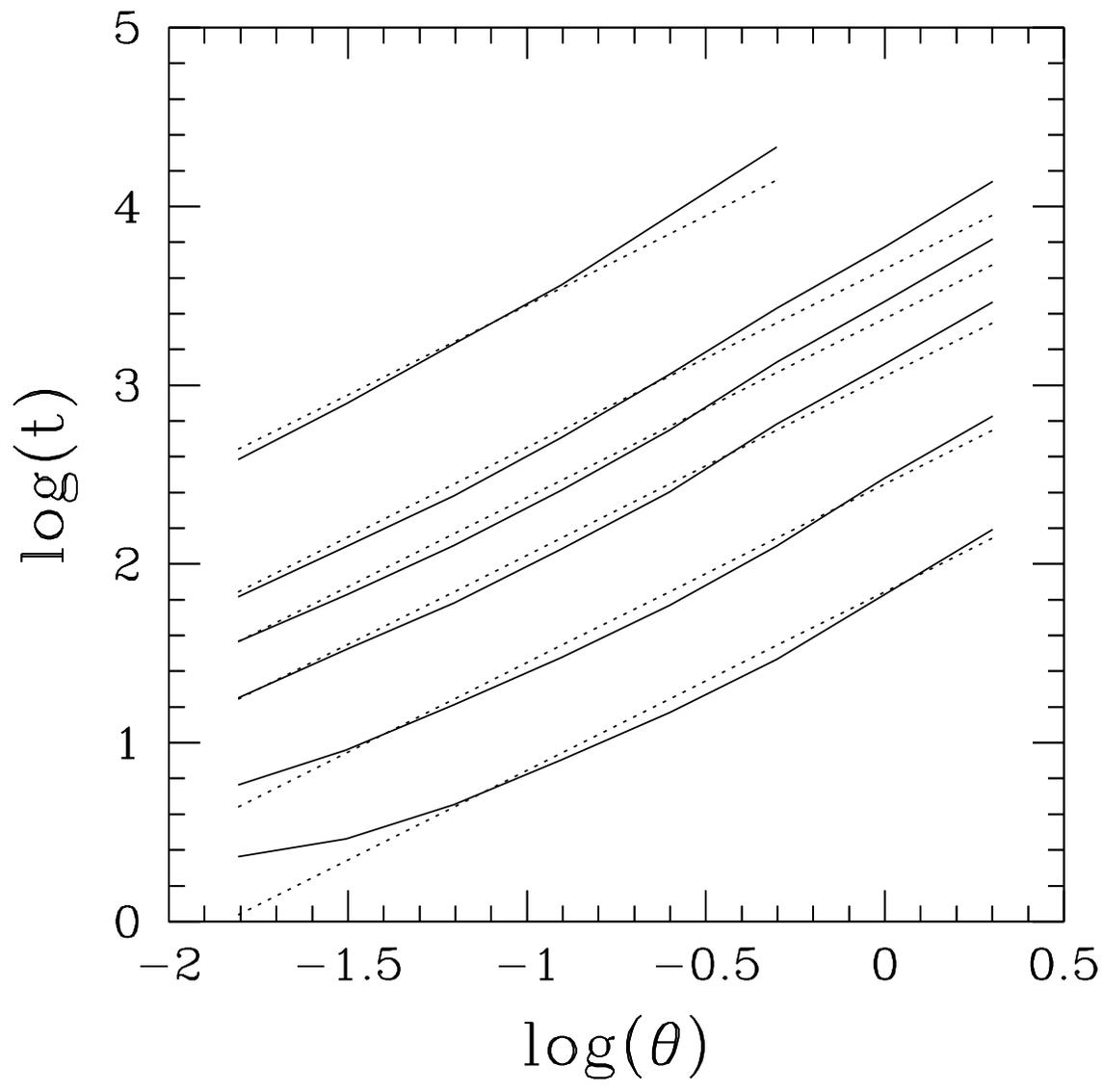
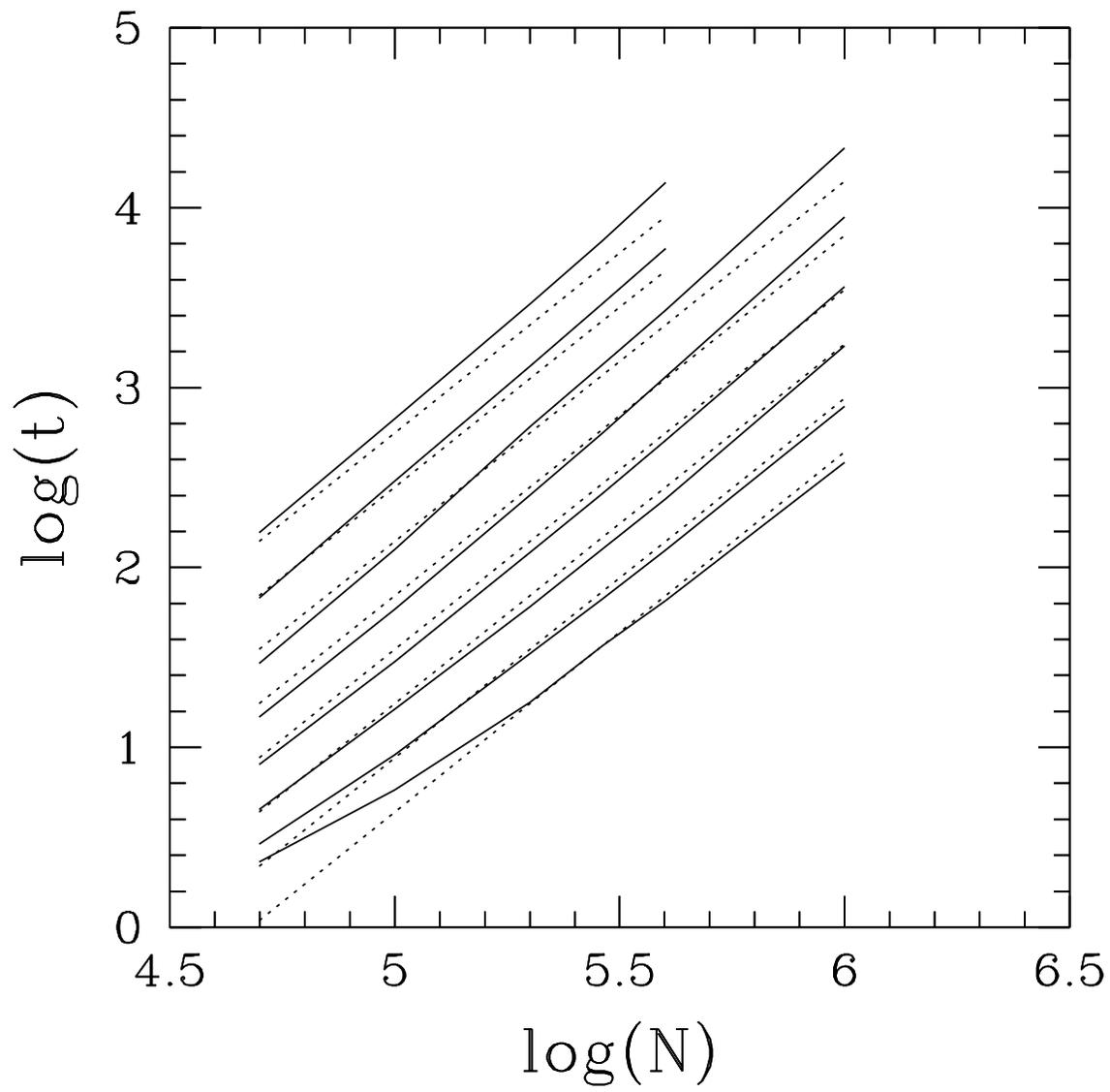
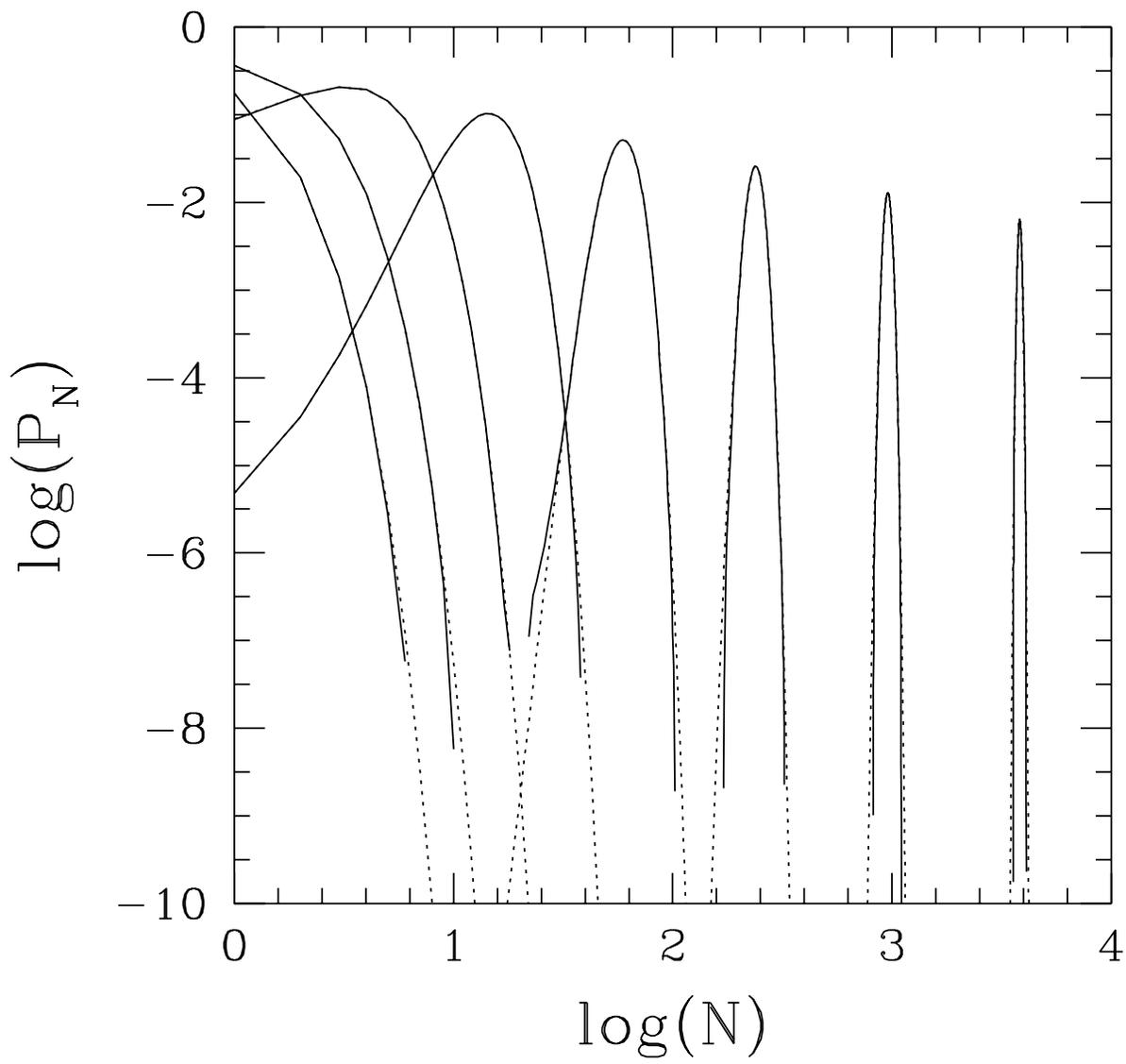
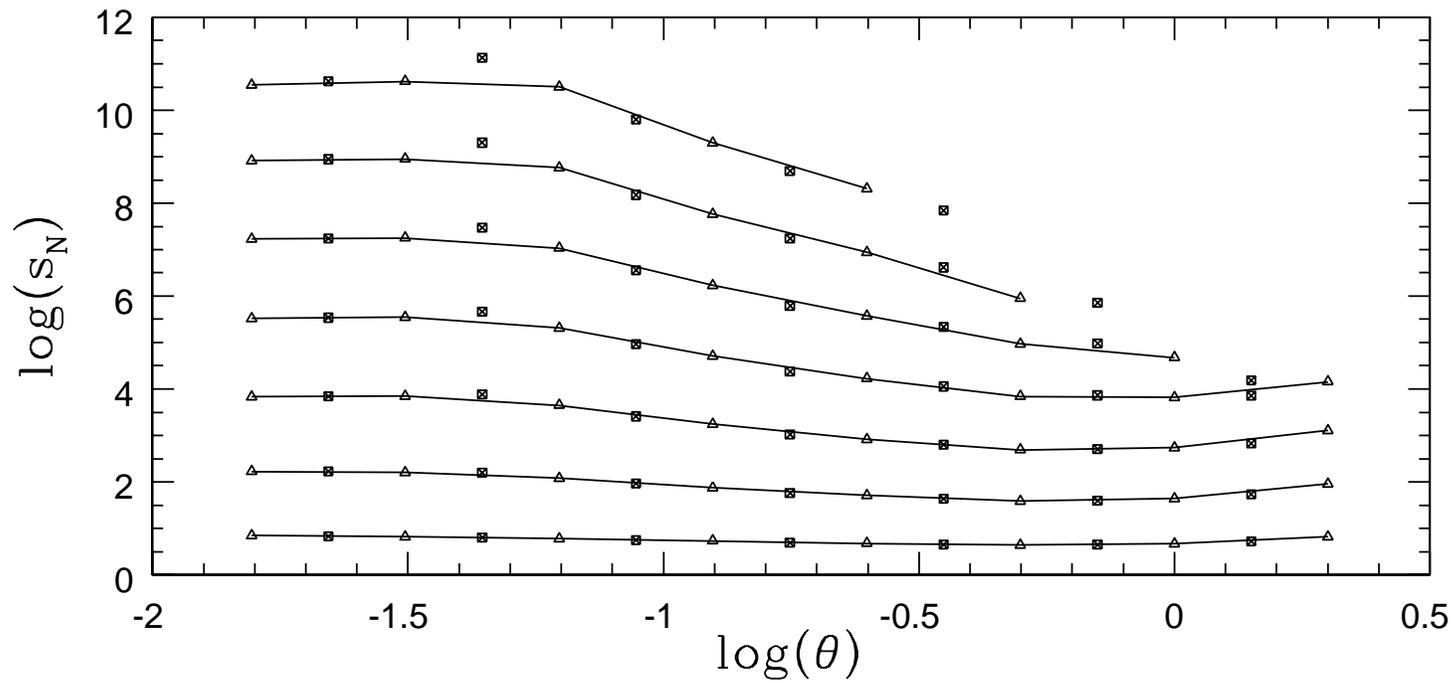
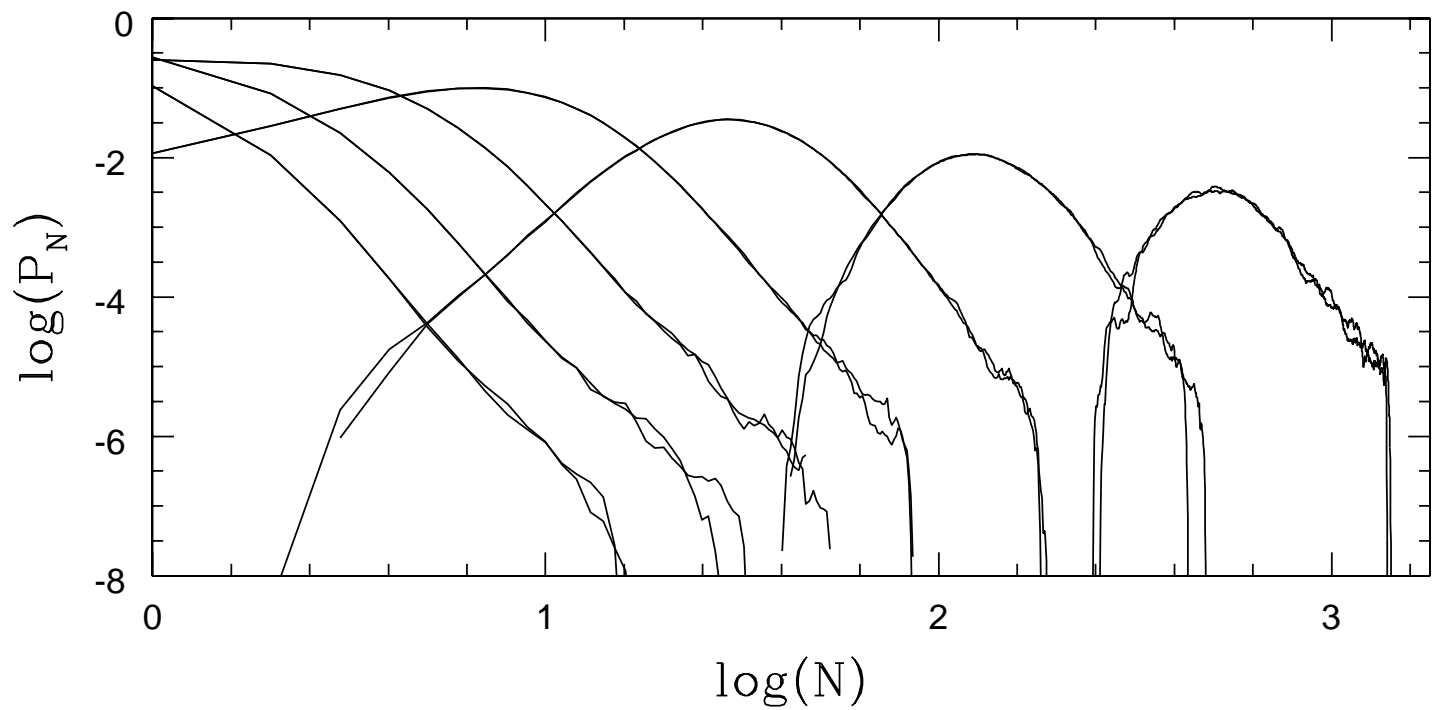
\end{document}